\documentclass[%
aps,prl,
floatfix,
preprint,
showpacs,preprintnumbers,nofootinbib,
amsmath,amssymb,
superscriptaddress]{revtex4-1}
\usepackage{graphicx}% Include figure files
\usepackage{dcolumn}% Align table columns on decimal point
\usepackage{bm}% bold math
\usepackage{hyperref}% add hypertext capabilities
\hypersetup{
	colorlinks=true,       % false: boxed links; true: colored links
	linkcolor=blue,          % color of internal links
	citecolor=blue,        % color of links to bibliography
}
\usepackage{color}
\usepackage{float}
\usepackage{amsmath}
\usepackage{autobreak}
\begin{document}
\title{Infinitely degenerate exact Ricci-flat solutions in $f(R)$ gravity}

\author{Semin Xavier}
 %   \email{seminxavier@iitb.ac.in}% Your name
    \affiliation{Department of Physics, Indian Institute of Technology, Mumbai, Maharashtra 400076}

\author{Jose Mathew}
%\email{josecherukara@gmail.com}
\affiliation{Department of Physics, The Cochin College, Kochi, Kerala, India }

\author{S. Shankaranarayanan}
%	\email{shanki@phy.iitb.ac.in}% Your name
    \affiliation{Department of Physics, Indian Institute of Technology, Mumbai, Maharashtra 400076}
     
\begin{abstract}
We obtain an infinite number of exact static, Ricci-flat spherically symmetric vacuum solutions for a class of $f(R)$ theories of gravity.  We analytically derive two exact vacuum black-hole solutions for the same class of $f(R)$ theories.  The two black-hole solutions have the event-horizon at the same point; however, their asymptotic features are different.  Our results suggest that no-hair theorem may not hold for generic modified gravity theories. We discuss the implications of our work to distinguish modified gravity theories from general relativity in gravitational wave detections.
\end{abstract}

\maketitle

Many unusual astrophysical phenomena occur near black-holes and neutron stars that are governed by the strong gravitational interactions~\cite{2014-Will-LRR,*2018-Will-Book}. Gravitational-wave observations of the merger of black-holes and neutron-stars have helped verify that General Relativity (GR) describes strong-gravity regimes~\cite{2016-Abbott.etal-PRL,*2019-Abbott.etal-PRDa}. These detections provide direct evidence of the existence of black-holes. The curvature singularities at the center of black-holes are unsatisfactory; paradoxically, GR predicts its failure to determine the future of the singularities~\cite{1976-Hawking-PRDa}.

In an effective field theory viewpoint, it is expected that the effective gravity action consists of the classical Einstein action plus a series of covariant, higher curvature terms that are important in the strong gravity regime~\cite{1977-Stelle-PRD,*1994-Donoghue-PRD,*2007-Burgess-ARNPS,*2011-Capozziello.DeLaurentis-PRep,*2012-Clifton.etal-PRep,*2017-Nojiri.etal-PRep,2010-Sotiriou.Faraoni-RMP,*2010-DeFelice.Tsujikawa-LRR,*2007-Woodard-Proc}. 
Also, in this regime, fields that are frozen at low energies may become dynamical and contribute to the dynamics. The detection of gravitational waves has provided a possibility to test modified gravity theories in the strong gravity regime~\cite{2013-Gair.etal-LRR,*2013-Yunes.Siemens-LRR,*2015-Nakano.etal-PRD,2011-Hild.etal-CQG,*2017-Abbott.etal-CQG}. 
Current constraints on deviations from GR are obtained by using parameterized post-Newtonian (PPN) formalism~\cite{1971-Thorne.Will-ApJ,*1971-Will-ApJ}. 
In this regime, PPN formalism is not sufficient, and there is no established formalism to test deviations from GR~\cite{2019-Barack.etal-CQG}. 

To circumvent this, recently, strong gravity diagnostic parameters are proposed to distinguish GR and modified gravity theories~\cite{2019-Shankaranarayanan-IJMPD,*2017-Bhattacharyya.Shankaranarayanan-PRD}. However, these parameters assume that the black-hole solutions in modified theories and GR are identical. In GR, the \emph{no-hair property of isolated event-horizons} leads to the expectation that all astrophysical black-holes are Kerr black-holes characterized by their masses ($M$) and spins ($J$). This is because isolated black-holes do not radiate and are axisymmetric~\cite{2009-Wiltshire.etal-Book,*2012-Chrusciel.etal-LRR,*2015-Teukolsky-CQG}. Despite a wealth of observational evidence, however, there is still no definite proof for their existence. The event horizon telescope is expected to directly image the black-hole shadow at the center of our Galaxy, and set constraints on deviations from Kerr~\cite{2014-Broderick.etal-ApJ,*2016-Johannsen.etal-PRL,*2019-Psaltis-GRG}.

In GR, Birkhoff's theorem guarantees that the general spherically symmetric vacuum solution is always static Schwarzschild metric~\cite{1973-Barnes-CMP,*2006-VojeJohansen-Ravndal-GRG,*2013-Schmidt-GRG}. This is because a spherically symmetric system cannot couple to higher spin excitations when spin-$0$ is absent
~\cite{1973-Misner.etal-Gravitation,*Wald:1984-bk}. However, a rotating star can have gravitational multipoles that are not the same as Kerr~\cite{2015-Teukolsky-CQG}.  Hence, in GR, 
Schwarzschild solution describes the space-time outside a non-rotating star.
 
{In this work, we explicitly obtain an infinite number of spherically symmetric vacuum solutions in $f(R)$ theories.} We obtain two black-hole solutions with an event-horizon at the same point. In GR, the zero-spin ($J \to 0$) limit of Kerr black-hole uniquely leads to the Schwarzschild solution. Thus, if there exists a large number of spherically symmetric vacuum solutions in $f(R)$, our results suggest that no-hair theorem also \emph{may not} hold for a class of $f(R)$ theories.

$f(R)$ theories of gravity are the straightforward modifications to GR~\cite{2010-Sotiriou.Faraoni-RMP}. The higher-order Ricci scalar terms encapsulate high-energy modifications to GR. Although the equations of motion are higher-order, they do not suffer from Ostr\"ogradsky instability~\cite{2007-Woodard-Proc}. Thus, $f(R)$ theories provide a natural arena for understanding many exhaustive features of strong-gravity. Unlike GR, $f(R)$ gravity has 11 dynamical variables --- 10 metric variables ($g_{\mu\nu}$) and the Ricci scalar ($R$). In other words, in $f(R)$ theories, the scalar curvature $R$, plays a non-trivial role in determining the metric itself~\cite{2019-Shankaranarayanan-IJMPD,1993-Hamity.Barraco-GRG,*2016-Tian-GRG}.

Although Schwarzschild black-hole is \emph{a solution} to vacuum $f(R)$ theories~\cite{2009-Cruz-Dombriz.etal-PRD,*2011-Moon.etal-GRG,*2015-Canate.etal-CQG}, it is unclear whether Schwarzschild is the unique vacuum solution. The reason for such a possibility to arise is that the new field equation is satisfied by $R$.  We show that $f(R)$ admits multiple space-time geometries with the horizon for the same stress-tensor configuration (vacuum in this case) {  \emph{without} transforming to conformal frame~\cite{2011-Sotiriou.Faraoni-PRL,*2010-Sebastiani.Zerbini-EPJC,*2011-Bergliaffa.etal-PRD,*2016-Gao.Shen-GRG,*2016-Amirabi.etal-EPJC,*2018-Calza.etal-EPJC}. Several studies have pointed physical nonequivalence of the Jordan and Einstein frames~\cite{Briscese:2006xu,*Capozziello:2006dj,*Capozziello:2010sc,*2015-Cognola.etal-PRD}.} The concept of the horizon is, in general, observer-dependent. However, for spherically symmetric static space-times, by \emph{horizon}, we refer to a horizon associated with static observers. An overprime denotes the derivative w.r.t $r$, overdot denotes the partial derivative w.r.t Ricci scalar ($R$), and $\kappa^2 = 8 \pi G/c^4$. 

\noindent \emph{\underline{The $f(R)$ model:}} 
The \emph{vacuum} $f(R)$ action is~\cite{2010-Sotiriou.Faraoni-RMP}:
\begin{equation}\label{Intreq1}
S[g_{\mu \nu}]=\frac{1}{2 \kappa^4 }\int d^{4} x \sqrt{-g}  \, f(R)
\end{equation}
where $f(R)$ is an arbitrary, smooth function of the Ricci scalar $R$. The modified Einstein tensor $({\cal G}_{\mu\nu})$ vanishes:
{\small
\begin{equation}
\label{eq:ModEinsEq}
{\cal G}_{\mu \nu}  \equiv  \left( R_{\mu \nu} - 
\nabla_{\mu} \nabla_{\nu} \right) \dot{f}(R) + g_{\mu \nu} \square \dot{f}(R)
-\frac{f(R)}{2} g_{\mu \nu}   = 0\, ,
\end{equation}
}
where $\dot{f}(R) \equiv F(R) = \partial f/\partial R$ and $\square=\nabla^{\mu} \nabla_{\mu}$. The generalized Bianchi identity leads to~\cite{1993-Hamity.Barraco-GRG}:
\begin{equation}
\label{eq:BianchifR}
\ddot{f}(R)  R_{\mu\nu} \nabla^{\mu} R  = 0 \, . 
\end{equation}
For GR, $\ddot{f}(R)  = 0$, and the above equation is trivially satisfied. However, $\ddot{f}(R)$ is non-zero for modified gravity theories. Hence, the generalized Bianchi identity (\ref{eq:BianchifR}) leads to four constraints on the Ricci tensor. While, GR and $f(R)$ have four constraints on the field variables, the number of dynamical variables is different. For the $f(R)$ action \eqref{Intreq1},  the trace of the field equation (\ref{eq:ModEinsEq}) is dynamical:
\begin{equation}
\label{eq:Trace}
R \, F(R) +3 \, \square F(R)  -2 \, f(R) =0
\end{equation}
{  As mentioned earlier, $f(R)$ gravity has 11 dynamical variables --- 10 metric variables ($g_{\mu\nu}$) and Ricci scalar ($R$).  Hence, the Ricci scalar plays a non-trivial role in determining the metric itself. }

One may still find the trivial solution where field equations reduce to the Einstein field equations with an effective cosmological constant and an effective gravitational constant~\cite{2015-Canate.etal-CQG}. This includes the case where $R=0$. Thus, all known black-hole solutions in GR also exist in this model.  Our interest in this work is to obtain \emph{non-trivial solutions}, taking into account the trace equation \eqref{eq:Trace}.

To model modified gravity in the strong-gravity regime, we consider $f(R)$ to be a polynomial in $R$:
\begin{equation}
\label{eqnfr1}
f(R)= %\sum_{i = 0}^{\infty} \beta_i \, R^i
\beta_{0} + \beta_{1} R+ \beta_{2} R^{2}+\cdots  + \beta_i \, R^i + \cdots\, ,
\end{equation}
where $\beta_{i}$'s are constants with appropriate dimensions. To keep the calculations tractable, we assume that $f(R)$ can be written in a binomial form:
\begin{equation}
\label{eq:modelfR}
f(R)= (\alpha_{0} + \alpha_{1} R)^{p} \quad \mbox{where} \quad 
\alpha_0, \alpha_1 > 0
\end{equation}
where $p$ is the power index. Thus, all the $\beta_i$'s in \eqref{eqnfr1} are related to the two constants $\alpha_{0}$ and $\alpha_{1}$. [$\alpha_0$ is dimensionless and $\alpha_1$ has dimensions of $[L]^{2}$.] 

For $p = 1$, we have: $f(R)= \alpha_{0} + \alpha_{1} R$. Physically, 
$\alpha_{0}$ acts like the cosmological constant and  
$\alpha_{1}$ modifies the Newton's constant. Since, we are interested in the strong-gravity corrections to GR, we take $p > 1$. In principle, $p$ need not be an integer. We now obtain 
generic, static spherically symmetric solutions for the above $f(R)$ model \emph{without} transforming to conformal frame~\cite{2011-Sotiriou.Faraoni-PRL}.

\noindent \emph{\underline{A class of exact solutions}:} 
The static, spherically symmetric metric in 4-D can be 
written in the following form:
\begin{equation}\label{metric1}
 ds^{2}=-A(r) e^{\delta(r)}  d t^{2}+\frac{d r^{2}}{A(r)}+
 r^{2} \left( d \theta^{2}+ \sin^{2} \theta \,  d \phi^{2} \right)
\end{equation}
where $A(r)$ and $\delta(r)$ are unknown functions of the Schwarzschild 
radial coordinate $r$. Substituting the above line-element in the modified Einstein's equations \eqref{eq:ModEinsEq} for the model  \eqref{eq:modelfR}, leads to the following %three field 
equations:
\begin{subequations}
\label{eq:ModEinsEq-Sph}
\begin{eqnarray}
\label{eq:ModEinsEq-T3}
{\cal G}^{t}_{t} &\equiv & T_{3}[A(r),\delta(r)]=0 \\
\label{eq:ModEinsEq-T4}
{\cal G}^{r}_{r} & \equiv & T_{4}[A(r),\delta(r)]=0 \\
\label{eq:ModEinsEq-T5}
{\cal  G}^{\theta}_{\theta} = {\cal G}^{\phi}_{\phi} &\equiv& T_{5}[A(r),\delta(r)]=0
\end{eqnarray}
\end{subequations}
where $T_{3}$, $T_{4}$ and $T_{5}$ are functions of 
$A(r)$ and $\delta(r)$, and their derivatives. More specifically, 
(i) $T_{3}$ and $T_{5}$ are non-linear, and \emph{contain up to 4th order derivatives} of $A(r)$ and $\delta(r)$, and
(ii) $T_{4}$ is non-linear and \emph{contain up to 3rd order derivatives} of $A(r)$ and $\delta(r)$. 
(iii) Even in the special case of $\delta(r) = 0$, ${\cal G}^{t}_{t} \neq {\cal G}^{r}_{r}$. Hence, we do not expect to get identical solutions as in GR.

\begin{figure}[!htb]
%\vspace*{-20pt}
    \centering
     \includegraphics[width=1.0\linewidth]{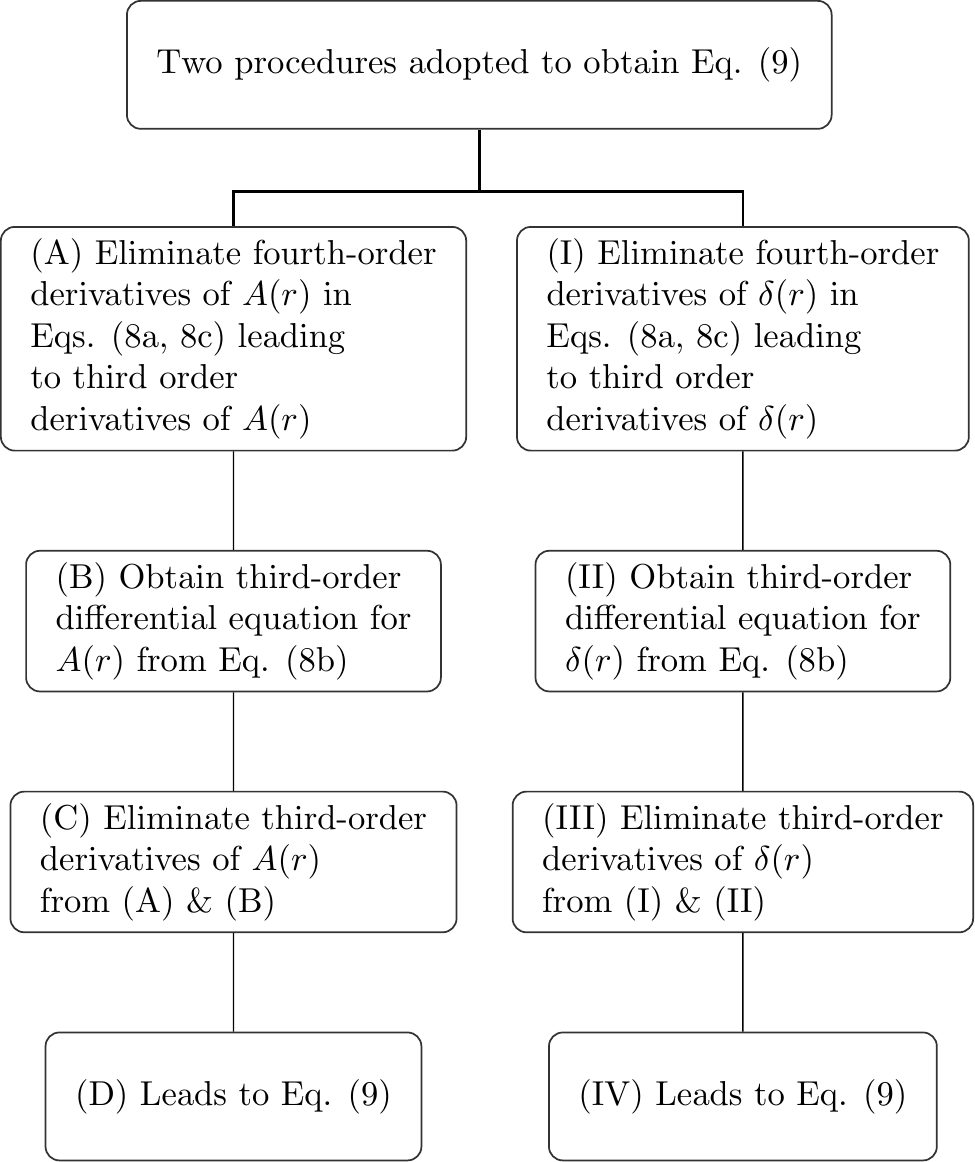}
\caption{Flow-chart of the two procedures leading to Eq. \eqref{eq:ReduEOM}.} 
\label{image}
\end{figure}

The exact forms of $T_3, T_4$ and $T_5$ are not relevant for the rest of the calculations; hence, they are not reported here.  They can be seen in the MAPLE code available in the \href{https://www.dropbox.com/sh/ca3nixx9weld57r/AADkzbbIQloHfFeV_2rVO2hca?dl=0}{Dropbox folder}. 
As expected, the equations of motion contain up to fourth-order derivatives in $A(r)$ and $\delta(r)$. Thus, an exact solution to these equations will contain up to four independent constants.  Fig. \eqref{image} contains the 
procedure we have adopted to reduce these equations 
into a product of two second-order  differentials in $A(r)$ and 
$\delta(r)$. Interestingly, both the procedures lead to:
% the following equation:
%
\begin{equation}
\frac{2}{r} \, \frac{(p-\frac{1}{2})}{p \, (p-1)} \, 
\frac{{T}_{1}[A(r),\delta(r)]}{(\Phi(r) + 4)} 
\frac{{T}_{2}[A(r),\delta(r)]}{(\Phi(r) - 2)}  = 0 
\label{eq:ReduEOM}
\end{equation}
where $\Phi(r) = r \left(\delta^{'}(r) + [\ln A(r) ]^{'}\right)$, and 
\vspace{-5pt}

\begin{widetext}
%\begin{subequations}
%\label{eq:defT1T2}
\begin{eqnarray}
\!\!\!\!\! {T}_{1}[A(r),\delta(r)] &=&
\left[ \Phi(r)+\frac{(\, p+1 \, )}{(\, p-\frac{1}{2}\,)} \,\right]
\left[\frac{\Phi(r)}{r}\right]^{'}  \!\!
%\left(\, \delta(r)^{''}+(\, \ln A(r)  \, )^{''} \right)
-\frac{3\,r}{2}\left[\, \delta'(r)^{3}+(\,[\,\ln A(r) \,]^{'} )^3 \,\right] 
\nonumber \\ 
& +&  \frac{ (3 \, \Phi(r) - 4) A'(r)}{r \, A(r)}  + \left(\,  2 \Phi(r) +1 \, \right)  \delta'(r) ^2   + + \frac{ (5\, \Phi(r) -2) A'(r)^2 }{2 A^2(r)}   \nonumber \\
\label{eq:defT1}
& + & 
 \frac{(\, A(r)+1\,)(\, p-1 \,) \, \Phi(r) }{(\, p - \frac{1}{2}) r^2 \, A(r)} 
- 4 \frac{( A(r)-1 \, )}{ r^2 \, A(r)}   
+ \frac{\alpha_{0} (\Phi(r) - 2 )}{2 \, \alpha_{1} \, A(r) \,(\, p - \frac{1}{2}) } 
\\
\label{eq:defT2}
T_{2}[A(r),\delta(r)] &=& 
r^2 \,A(r) \Big[ \left(\frac{\Phi(r)}{r}\right)^{'}+\left[ \left(\frac{4+r}{2r}\right) +   \frac{3}{2} [\ln(A(r)) ]^{'}  \right]\left(\frac{\Phi(r)}{r}\right) \Big] 
%\nonumber 
\\
%\end{eqnarray}
%\begin{eqnarray}
&-& \frac{r^2 \,A(r)}{2} \Big( [ \ln(A(r))']^{2}  + \,\left( \frac{4+r}{r}\right) [ \ln(A(r))]^{'} - \frac{4}{r} \,\left[ \frac{1}{r} +2\right] \Big)
- 2  \left(  1+ \frac{\alpha_{0} \, r^2}{2 \, \alpha_{1}}\right) \, . \nonumber
\end{eqnarray} 
\end{widetext}

This is the first important result regarding which we would like to stress the following points: First, as mentioned above, we have obtained the same equation \eqref{eq:ReduEOM} using two different approaches. This implies that Eq. \eqref{eq:ReduEOM} is a unique differential equation for this $f(R)$ model for the static spherically symmetric space-time \eqref{metric1}. Second, the above simplified equation is a product of two second-order non-linear differentials of 
$A(r)$ and $\delta(r)$.  Thus, the above equation drastically simplifies the procedure to obtain the exact black-hole solutions 
for any value of $p$. Third, the immediate consequences of the above equation are the conditions it imposes on $p$, $A(r)$ and $\delta(r)$. More specifically, demanding a non-trival solution to be satisfied for any finite value of $r$, 
leads to:
%we get,
%
\begin{equation}
\label{eq:constraints}
p \neq 0 \, , ~1/2~ ,~1 \, ;  \, \Phi(r) \neq -4 ~ \mbox{or} ~  2 \, .
\end{equation} 
Since $p = 1$ is not allowed, the non-trivial exact solutions are \emph{only valid} for modified theories. While $p = 0$ and $1$ will lead to divergence, $p = 1/2$ will lead to trivial solutions. Fourth, the above condition on $\Phi(r)$ implies
\begin{equation}
\delta(r) + \ln A(r) \neq \ln(r^2) ~ \mbox{or} ~ \ln(r^{-4})
\end{equation}
If we assume $\delta(r) = {\rm constant}$, then $A(r) \neq c_0 r^2 + c_1 r^{-4}$, where $c_0, c_1$ are constants.  Lastly, non-trivial solutions for the above equation \eqref{eq:ReduEOM} are possible if $T_1$ or $T_2$  vanish:
\begin{equation}
\label{eq:constraints02}
T_{1}[A(r),\delta(r)] = 0 \qquad \mbox{or} \qquad 
T_{2}[A(r),\delta(r)] = 0
\end{equation}
In principle, for a given $A(r)$, we can have two forms of $\delta(r)$ that 
satisfy either $T_1$ or $T_2$ vanish.  This leads to the question: how to obtain $A(r)$? Using Eq.~\eqref{eq:constraints}, we get
\begin{equation}\label{eqnadt}
\delta'(r) + ( \ln[A(r)] )^{'}  = \mu(r) \, ; \quad 
\mu(r) \neq \frac{-4}{r}~~\mbox{or}~~\frac{2}{r}
\end{equation}
For a given $\mu(r)$, we have a functional relation between $A(r)$ and $\delta(r)$. Substituting this relation in the constraint  \eqref{eq:constraints02}, we obtain a differential equation in terms of $A(r)$ or $\delta(r)$. Note that $\delta(r)=0$ \emph{trivially} satisfies Eqs. \eqref{eq:defT1}.

{  Thus, for the $f(R)$ model \eqref{eq:modelfR}, two branches of solutions exist: $T_1 = 0$ or $T_2 = 0$.} Since, $\mu(r)$ is arbitrary, we can obtain \emph{infinite vacuum  solutions} for the spherically symmetric metric \eqref{metric1}. This leads to the question: Whether any arbitrary function $\mu(r)$ satisfying $T_1 = 0$ or $T_2 = 0$ is a solution to the field equations \eqref{eq:ModEinsEq-Sph}?  To address this, %for any $p$, 
we write a formal solution to Eq.~\eqref{eqnadt} as
\begin{equation}
\label{eq:T1cond-Aform}
A(r)=e^{-\delta(r)} \gamma(r) ~ \mbox{where} ~
\gamma(r) = 
\exp \left( \int \mu(r) dr \right) .
\end{equation}
Substituting $A(r)$ in-terms of $\delta(r)$ in $T_{2}[A(r),\delta(r)] = 0$, we 
obtain a differential equation in $\delta'(r)$. Substituting these in Eq.~\eqref{eq:ModEinsEq-Sph}, we get, $T_{3}[\delta(r),\gamma(r)]=
T_{5}[\delta(r),\gamma(r)]$, 
%
%\begin{subequations}
\begin{eqnarray}
\label{eq:ModEinsEq-Sph02}
T_{3} &=&
4 \alpha_{1}  p (p-1) (p-2) \gamma^7(r) e^{-\delta(r)} 
\left[\delta'(r) r-8\right]^2 \,  T_{2}^2  \\
T_{4} &=& 
-2 \alpha_{1} p (p-1) \gamma^5(r) e^{-\delta(r)} 
\left[ \delta'(r) r-8\right] 
\left[r \ln [\gamma(r) ]^{\prime} +4  \right]  
T_{2} \nonumber
\end{eqnarray}
%\end{subequations}
%
Since, we have obtained the above expressions using the condition $T_{2}[A(r),\delta(r)] = 0$, we have $ {\cal G}^{t}_{t}={\cal G}^{r}_{r} = 
{\cal G}^{\theta}_{\theta}={\cal G}^{\phi}_{\phi}=0$. Thus, $A(r)$ obtained in Eq.~\eqref{eq:T1cond-Aform} satisfying  $T_{2}[A(r),\delta(r)] = 0$ 
is an exact solution for the $f(R)$ model \eqref{eq:modelfR}. Since $A(r)$ depends on the arbitrary function $\mu(r)$, for the same observer with Schwarzschild time $t$, there exists {  
\emph{infinite number} of exact static, spherically symmetric solutions satisfying Ricci-flat condition $R = -\alpha_0/\alpha_1$. }

This is the crucial result of this work. As mentioned earlier, the Birkhoff theorem in GR guarantees that the most general spherically symmetric vacuum solution is the static Schwarzschild solution~\cite{1973-Barnes-CMP,*2013-Schmidt-GRG}. However, in $f(R)$ gravity, the trace equation \eqref{eq:Trace} provides a non-trivial structure for the Ricci scalar as a function of $r$ leading to an infinite set of static solutions for $f(R)$ theories of gravity. {The non-existence of Birkhoff's theorem in $f(R)$ theories is known for sometime~\cite{1984-Riegert-PRL}, recently, several authors have tried to confirm/infirm 
Birkhoff's theorem in the conformal frame~\cite{2011-Sotiriou.Faraoni-PRL,2011-Oliva.Ray-CQG,*2012-Capozziello.Saez-Gomez-AnP}. 
Here, we have not made any approximation or performed a conformal transformation to obtain exact solutions.} Our results confirm that the Birkhoff theorem is not valid for $f(R)$ gravity theories. Our analysis and results are valid even if the conformal transformations to the Einstein frame is not well-defined. 

It is also important to compare our results with that of Jaime et al~\cite{2010-Jaime.etal-PRD}: Jaime et al considered $f(R)$ models that satisfy two conditions: $\partial f(R)/{\partial R} >0$ and $\partial^2 f(R)/{\partial R^2}  > 0$. This is because the authors' focus was to obtain solutions of relativistic extended objects with external matter fields and not vacuum black-hole solutions. However, the black-hole solutions we obtain do not satisfy these conditions. 

Note that, unlike $T_2$, $T_1$ is not a common factor of the field equations \eqref{eq:ModEinsEq-Sph02}; hence, $T_1=0$ alone can not provide valid solutions. Thus, beside $T_1 = 0$, we must use either of the three equations \eqref{eq:ModEinsEq-Sph} to obtain a solution. We now obtain two particular vacuum, Ricci-flat black-hole solutions where $A(r)$ is the same, with different $\delta(r)$. 

\noindent \emph{\underline{Two black-hole solutions}:}~{  As noted above, $\delta(r) = 0$ trivially satisfies field equations \eqref{eq:ModEinsEq-Sph}. Substituting $\delta(r) = 0$ in Eq. \eqref{eq:defT2} leads to:}
\begin{equation} 
\label{eqnA2}
A(r)=1+C_{2}  r^{2} - \frac{C_{3}}{r^{2}} \quad \mbox{where}~~ 
C_{2}=\frac{\alpha_{0}}{12 \alpha_{1}} > 0
\end{equation}
which satisfies the null-energy condition~\cite{1973-Misner.etal-Gravitation}.
$C_3$ is a constant of integration. We like to list the following important points regarding this solution: First, it is easy to verify that the above solution satisfies the modified Einstein's equations \eqref{eq:ModEinsEq-Sph}. Second,  
physically, $C_2$ acts like an effective cosmological constant. For $C_3 > 0$, the metric \eqref{metric1} has a horizon at 
 \begin{equation}
 \label{eq:Sol1-horizon}
 r_{h}^2= (\sqrt{1 + 4 C_2 C_3} - 1)/({2 C_2})
 \end{equation} 
In the limit of $\alpha_0 \to 0$, $C_2 \to 0$, the metric \eqref{metric1} has a horizon at $r = \sqrt{C_3}$. Thus, $\alpha_0 \to 0$ is a smooth limit. Third, the term $c_3/r^2$ is reminiscent of the charge in the Reissner-Nordstr\"om solution in GR~\cite{1973-Misner.etal-Gravitation}.  In GR, $C_3/r^2$ term can not exist without the mass ($1/r$) term. 
In our case, the $1/r^2$ term is present in the absence of $1/r$ term. This result is similar to the one obtained sometime back in the context of black-holes on the brane~\cite{2000-Dadhich.etal-PLB}. Physically, $C_3$ corresponds to the black-hole mass. The Kretschmann scalar for this solution is:
\begin{equation}
R^{ \alpha \beta \gamma \delta } R_{\alpha \beta \gamma \delta }= 
24 \, C_2^2 + 56 C_{3}^{2}/{r^{8}}.
\end{equation}
Thus, the metric has a singularity at $r=0$ and is finite everywhere else.
For finite $\alpha_0$, the Kretschmann scalar is positive at asymptotic infinity which corresponds to asymptotic de Sitter space-times in GR~\cite{1973-Misner.etal-Gravitation}. {  Similarly, $R^{\alpha \beta \gamma \mu }R_{\alpha \gamma} R_{\beta \mu}$ is singular at $r=0$ and is finite everywhere else. 
This implies that the spherically symmetric solution \eqref{eqnA2} is a black-hole solution with an event-horizon at $r_h$ \eqref{eq:Sol1-horizon}.}  Fourth, the above solution is a particular case of a general solution that satisfies the null-energy condition. We have provided a general solution in the Appendix.  

{  We obtained \eqref{eqnA2} for $\delta(r) = 0$. As noted above, this is one particular choice of many choices allowed in Eq. \eqref{eq:T1cond-Aform} satisfying $R = -\alpha_0/\alpha_1$. We now exercise this freedom and substitute the above form of $A(r)$ in $T_{2}[A(r),\delta(r)] = 0$. Substituting \eqref{eqnA2} in \eqref{eq:defT2}, we get:}
\begin{equation}
\delta''(r) + \frac{1}{2}\delta'(r)^2 + \frac{(5 C_2 r^4 + 2 r^2 + C_3)}{(C_2 r^4 + r^2 - C_3)} \frac{\delta'(r)}{r} = 0
\end{equation}
where $C_2$ is defined in Eq. \eqref{eqnA2}. The above differential equation has the following exact solution:
\begin{equation}\label{eqndelap1}
e^\frac{\delta(r)}{2} = C_4 
+ \frac{C_5}{\sqrt{A(r)}} \left(2 C_2  r+ \frac{1}{r} \right)
%{C_5} \frac{(2 C_2 r^2  + 1)}{(4 C_2 C_3 + 1)}
%\frac{1}{(C_2 r^4 + r^2 - C_3)^{1/2} }
%%\frac {( [ 24{\it C_3}\alpha_1 \alpha_0-72{\alpha_1}^{2}](\sqrt{\alpha_0{r}^{4}+12\alpha_1{r}^{2}+12{\it C_3}\alpha_1}){\it C_4}+[\alpha_0{r}^{2}+6\alpha_1] {\it C_5} )^{2}}{2304{\alpha_1}^{2} ( {\it C_3}\alpha_0-3 \alpha_1 )^{2} (\alpha_0 {r}^{4}+12 \alpha_1{r}^{2}+12{\it C_3}\alpha_1 )} 
\end{equation}
where $C_{4}$ and $C_5$ are arbitrary constants. 
{  Demanding that the metric coefficients are real for all values of $r$ leads to $C_4 = 0$, and $C_5$ can take any real number. The event-horizon for this solution is also given by \eqref{eq:Sol1-horizon}. 
For this solution also, Kretschmann scalar and $R^{\alpha \beta \gamma \mu }R_{\alpha \gamma} R_{\beta \mu}$ are only singular at $r=0$ and finite everywhere else. The surface gravity of the horizon is zero. 
%(See Secs. I and II of Supplementary). 
QNMs arising from these black-holes will not satisfy isospectral relations~\cite{2019-Shankaranarayanan-IJMPD}.}

We have reconfirmed the results in the earlier part and shown that at least two vacuum black-hole solutions exist for $f(R)$ model \eqref{eq:modelfR}. To our knowledge, this is a new result for any modified theories of gravity and confirms that the Birkhoff's theorem is not valid for $f(R)$ theories. We can also understand the results from the difference in the order of the equations of motion in GR and $f(R)$. In GR, the equations of motions are second-order and hence, can have a maximum of two integration constants. However, in $f(R)$, the equations of motion \eqref{eq:ModEinsEq} are fourth-order, hence, can have a maximum of four arbitrary constants ($C_2, C_3, C_4, C_5$).  

Our analysis and results are valid even if the conformal transformation to the Einstein frame is not well-defined. The key ingredient in the proof of the Birkhoff theorem in GR is the absence of spin-$0$ modes in the linearized field equations. The spherically symmetric space-time cannot couple to higher-spin excitations when spin-$0$ is absent~\cite{1973-Misner.etal-Gravitation,1984-Riegert-PRL}. In the case of $f(R)$ theories, the differential equation satisfied by the Ricci scalar plays the role of spin-$0$ modes. Thus, a non-trivial dependence between the metric and the Ricci scalar, \emph{in general}, leads to the breaking of the Birkhoff theorem in $f(R)$. 

It is important to note that the two black-hole solutions we have obtained are among the infinite number of exact static, Ricci-flat spherically symmetric vacuum solutions for the $f(R)$ model~\eqref{eq:modelfR}. 
To obtain regular solutions, we need to impose boundary conditions at a finite $r$, which can be related to $\mu(r)$.  As mentioned earlier, through $\mu(r)$, the functional form of $A(r)$ and $\delta(r)$ determine the solutions' characteristics. Such solutions with rotation can help to describe the gravitational field outside the Neutron star. This is currently under investigation. 

\noindent \emph{\underline{Conclusions and Discussions:}} 
We obtained an infinite number of exact static spherically symmetric vacuum solutions for $f(R)$ gravity. To emphasize this feature, we obtained two exact vacuum black-hole solutions with an event-horizon at the same point with different asymptotic features. {Our results confirm that the Birkhoff theorem is not valid for all modified gravity theories.} In GR, the zero-spin ($J \to 0$) limit of Kerr black-hole uniquely leads to the Schwarzschild solution. Thus, if there exists a large number of spherically symmetric vacuum solutions in $f(R)$, our results suggest that no-hair theorem also \emph{may not} hold for $f(R)$ theories.
Our analysis is a step to infirm/confirm the no-hair theorem in $f(R)$ theory.
One possible way to obtain an axial solution is to use  Newman-Janis algorithm~\cite{1965-Newman.Janis-JMP,*1998-Drake-Szekeres-GRG}. This is currently under investigation.

Unlike in the literature, we have obtained the exact solutions \emph{without} transforming to a conformal frame. It is then natural to ask {  what happens in the conformal frame? }
Under conformal transformations ($\tilde{g}_{\mu\nu} =  F(R) \, g_{\mu\nu}$), the action \eqref{Intreq1} transforms to~\cite{2010-Sotiriou.Faraoni-RMP}:
\begin{equation}
    S^{E}=\int d^4x\sqrt{-\tilde{g}}\left[\frac{1}{2\kappa^2}\tilde{R}-\frac{1}{2}\partial^{\alpha} {\varphi}\partial_{\alpha}{\varphi}-V({\varphi})\right]
\end{equation}
{ 
For our model \eqref{eq:modelfR}, ${\varphi},$ and $V(\varphi)$ are given by:
\begin{eqnarray}
\sqrt{\frac{2 \kappa^2}{3}}  \varphi &=&  
%(p - 1)  \ln \left(\frac{T_2[A(r), \delta(r)]}{r^2} \right) + \ln(\alpha_1 p) =
  (p - 1) \ln (\alpha_0 + \alpha_1 R) + \ln(\alpha_1 p) \\
V(\varphi) &=& \frac{1}{2 \kappa^2 p^2 \alpha_1} 
\frac{(p - 1) \alpha_1 R - \alpha_0}{(\alpha_0 + \alpha_1 R)^{p -1} } 
\end{eqnarray}
For all solutions with $R = -\alpha_0/\alpha_1$, hence,  $V(\varphi)$ diverges and the theory is ill-defined in the Einstein frame. Thus, such $f(R)$ models do not have an equivalent description in the Einstein frame. It is important to note that other authors have pointed out the nonequivalence of the Jordan and Einstein frames in other $f(R)$ models~\cite{Briscese:2006xu,*Capozziello:2006dj,*Capozziello:2010sc}. }

Our analysis shows the deficiency of finding solutions in the conformal frame. {  The conformal transformations are  ill-defined if the conformal factor vanishes.} However, the solution corresponding to another branch
$T_1[A(r), \delta(r)] = 0$ will be well-defined in the conformal frame. We 
plan to use the publically available NeuroDiffEq package to obtain new non-trivial solutions in $f(R)$ models~\cite{2020-Chen.etal-JOSS}.

To keep the calculations tractable, we have used a binomial form for $f(R)$. However, the solutions we have derived should be valid for any $f(R)$ model.
The condition that $F(R)$ vanishes ensures that all the field equations are satisfied when $R$ takes a constant value. We can build infinitely many interesting $f(R)$ models for the same metric, which yields a constant $R$. 

One of the prospects of the gravitational wave observations is to find signatures for the modified gravity theories. Suppose the modified theories belong to one of the degenerate classes with $T_2 = 0$. In that case, our analysis shows that the prospect of detection needs different methodologies than the one that is currently used~\cite{2019-Barack.etal-CQG}. 

\begin{acknowledgments}
The authors thank N. Dadhich, Saurya Das, Bala Iyer, Sayan Kar, K. Lochan, T. Padmanabhan, and A. Pai for discussions. This work was made possible by the use of \href{http://grtensor.phy.queensu.ca/}{GRTensor 3 package} for MAPLE.  SX is financially supported by the MHRD fellowship at IIT Bombay. The work is supported by the MATRICS SERB grant. %(MTR/2019/000077).
\end{acknowledgments}

\appendix
\section{Appendix: General class of solutions}
\label{sec:App}

In this Appendix, we obtain the form of $\delta(r)$ for two other forms of $A(r)$. Note that Eq. \eqref{eqnA2} is a particular case of 
Form 1 given below.
 
\begin{itemize}
    \item {\bf Form 1:} The general form of $A(r)$ that satisfies Eq. 11 for the case $\delta(r) = 0$ is:
\begin{equation}
    A(r)=1+C_2\,r^2-\frac{C_3}{r^2}+\frac{C_4}{r}
    \label{eqnR4}
\end{equation}
Substituting the above form of $A(r)$ in Eq. 11, we get:
\begin{equation}
    \delta(r)=2\ln  \left( \frac{1}{2}\,{C_0}\,\int \!{\frac {r}{ \left( -\alpha_{0}{r}^{4}-12
\,{C_4}\,\alpha_{1}\,r-12\,\alpha_{1}\,{r}^{2}+12\,{C_3}\,\alpha_{1}\, \right) ^{3/2}}}\,{\rm d}r +\frac{C_{5}}{2} \right)
\label{eqnR5}
\end{equation}
which is simplified to:
\begin{equation}
    e^{\delta(r)/2}={C_0}^{'}\,\int \!{\frac {r}{ \left( -\alpha_{0}{r}^{4}-12
\,{C_4}\,\alpha_{1}\,r-12\,\alpha_{1}\,{r}^{2}+12\,{C_3}\,\alpha_{1}\, \right) ^{3/2}}}\,{\rm d}r+C_{5}^{'} 
\label{eqnR6}
\end{equation}
where $C_{5}^{'}=\frac{C_{5}}{2}$ and $C_{0}^{'}=\frac{C_{0}}{2}$ are constants. 
\item {\bf Form 2}: Setting $C_3 = 0$ in Eq. \eqref{eqnR4}, we have:
    \begin{equation}
    A(r)=1+C_2\,r^2+\frac{C_4}{r}
    \label{eqnR1}
\end{equation}
Substituting the above form of $A(r)$ in Eq. 11, we get:
\begin{equation}
    e^{\delta(r)/2}={C_0}^{'}\,\int \!{\frac {1}{ \left( \alpha_{0}{r}^{3}+12
\,{C_4}\,\alpha_{1}+12\,\alpha_{1}\,r \right) ^{3/2}\sqrt {r}} }\,{\rm d}r+C_{5}^{'} 
\label{eqnR3}
\end{equation}
\end{itemize}
We would like to point the following regarding the above exact solutions: 
First, these are exact solutions to our $f(R)$ model \eqref{eq:modelfR}. These expressions correspond to valid black-hole solutions. Second, these solutions are consistent with the results obtained earlier in the literature (see Ref.~\cite{2008-Saffari.etal-PRD}). However, as can be seen, the above two forms of $\delta(r)$  can only be expressed in an integral form. 

%\bibliography{references}
%merlin.mbs apsrev4-1.bst 2010-07-25 4.21a (PWD, AO, DPC) hacked
%Control: key (0)
%Control: author (8) initials jnrlst
%Control: editor formatted (1) identically to author
%Control: production of article title (-1) disabled
%Control: page (0) single
%Control: year (1) truncated
%Control: production of eprint (0) enabled
%

\end{document}